\documentclass[sigconf,9pt]{acmart}
\graphicspath{{./figures/}{../}}

\usepackage{graphicx}
\usepackage{textcomp}
\usepackage{xcolor}
\usepackage{balance}
\usepackage{booktabs}
\usepackage{pifont}
\usepackage{pgfplots}
\usepackage{pgfplotstable}
\pgfplotsset{compat=1.8}
\usepackage{comment}
\usepackage{multirow}
\hypersetup{
    colorlinks = true,
    citecolor  = blue,
    linkcolor  = blue,
    urlcolor   = blue,
    pdftitle={MappingEvolve: LLM-Driven Code Evolution for Technology Mapping},
    pdfauthor={Rongliang Fu, Yi Liu, Qiang Xu, Tsung-Yi Ho},
    pdfsubject={Proceedings of the 63rd ACM/IEEE Design Automation Conference (DAC)}
}
\usepackage{resizegather}
\usepackage[subrefformat=parens,farskip=0pt,justification=centering]{subfig}

\usepackage[labelsep=period,font=small]{caption}
\usepackage{url}
\usepackage{threeparttable}
\usepackage{physunits}
\usepackage[round-pad=false, round-mode=places,round-precision=2,group-separator={,},output-decimal-marker={.}]{siunitx}

\usepackage{amsthm}

\newtheorem{definition}{Definition}
 
\makeatletter
\newif\if@restonecol
\makeatother

\usepackage[linesnumbered,ruled,vlined]{algorithm2e}
\usepackage{algpseudocode}
\usepackage{amsmath,amsfonts}
\SetKwRepeat{Do}{do}{while}
\SetAlgoInsideSkip{small}
\allowdisplaybreaks

\usepackage{amsthm}
\usepackage{cleveref}
\Crefformat{figure}{Fig.~#2#1#3}                           
\Crefname{subfigure}{Fig.}{Figs.}
\Crefname{figure}{Fig.}{Figs.}
\Crefformat{table}{TABLE~#2#1#3}                           
\Crefname{table}{TABLE}{TABLEs}

\usepackage{tikz}
\usetikzlibrary{positioning,arrows.meta}
\usepackage{enumitem}

\newcommand{\minisection}[1]{\vspace{.06in}\noindent{\textbf{#1}}.}

\iftrue
    \setlist{leftmargin=12pt}
\fi

\copyrightyear{2026}
\acmYear{2026}
\setcopyright{cc}
\setcctype{by-nc-nd}
\acmConference[DAC '26]{63rd ACM/IEEE Design Automation Conference}{July 26--29, 2026}{Long Beach, CA, USA}
\acmBooktitle{63rd ACM/IEEE Design Automation Conference (DAC '26), July 26--29, 2026, Long Beach, CA, USA}
\acmDOI{10.1145/3770743.3803988}
\acmISBN{979-8-4007-2254-7/2026/07}

\begin{document}

\title{MappingEvolve: LLM-Driven Code Evolution for Technology Mapping}

\author{Rongliang Fu}
\affiliation{%
  \fontsize{8}{10}\selectfont
  \institution{Chinese University of Hong Kong}%
  \city{Hong Kong}%
  \country{China}%
}

\author{Yi Liu}
\affiliation{%
  \fontsize{8}{10}\selectfont
  \institution{Chinese University of Hong Kong}%
  \city{Hong Kong}%
  \country{China}%
}

\author{Qiang Xu}
\affiliation{%
  \fontsize{8}{10}\selectfont
  \institution{Chinese University of Hong Kong}%
  \city{Hong Kong}%
  \country{China}%
}

\author{Tsung-Yi Ho}
\affiliation{%
  \fontsize{8}{10}\selectfont
  \institution{Chinese University of Hong Kong}%
  \city{Hong Kong}%
  \country{China}%
}

\renewcommand{\shortauthors}{Rongliang Fu et al.}

\begin{abstract}
  Technology mapping is a critical yet challenging stage in logic synthesis. While Large Language Models (LLMs) have been applied to generate optimization scripts, their potential for core algorithm enhancement remains untapped. We introduce MappingEvolve, an open-source framework that pioneers the use of LLMs to directly evolve technology mapping code. Our method abstracts the mapping process into distinct optimization operators and employs a hierarchical agent-based architecture, comprising a Planner, Evolver, and Evaluator, to guide the evolutionary search. This structured approach enables strategic and effective code modifications. Experiments show our method significantly outperforms direct evolution and strong baselines, achieving 10.04\% area reduction versus ABC and 7.93\% versus mockturtle, with 46.6\%--96.0\% $S_{overall}$ improvement on EPFL benchmarks, while explicitly navigating the area--delay trade-off. Our code and data are available at \url{https://github.com/Flians/MappingEvolve}.
\end{abstract}


\maketitle


\section{Introduction}
\label{sec:introduction}

Logic synthesis plays a pivotal role in the circuit design process, primarily comprising two sub-processes: logic optimization~\cite{lee2023peephole, dc_rewriting, fu2026chop, fu2026dclog, fu2026elogic} and technology mapping~\cite{ABC,calvino2022mapping}. Among these stages, technology mapping is particularly crucial as it bridges the gap between logical design and physical design.

Existing optimization methods for technology mapping primarily focus on three approaches:
i) cut selection to filter for logic-level or physical-level superior cuts, thereby obtaining better gate-level netlists, such as Priority Cuts~\cite{alan2007priority}, SLAP~\cite{neto2021slap}, LEAP~\cite{chigarapally2025LEAP}, and PigMAP~\cite{pan2025pigmap};
ii) multi-output or application-specific cell exploitation, such as emap~\cite{calvino2023emap} and dual-output LUT~\cite{shang2025dualoutput};
iii) standard cell library filtering and extension to shrink the search range or improve cell quality, such as MapTune~\cite{liu2025maptune} and TeMACLE~\cite{fu2025temacle}.
While these advances improve mapping quality, they predominantly refine heuristic parameters and search strategies rather than the core algorithmic logic within the mapper itself.

Recent advances in Large Language Models (LLMs), including GPT-5~\cite{gpt5}, DeepSeek-V3~\cite{deepseekv3}, and Qwen~\cite{qwen3}, have demonstrated remarkable capabilities in code comprehension and algorithmic reasoning. These capabilities have led to initial studies applying LLMs to logic synthesis, such as ChatLS~\cite{zheng2025chatls} and LLSM~\cite{huang2025llsm}. However, these studies mainly focus on generating optimization scripts rather than evolving the logic synthesis algorithms themselves.
Unlike conventional automated parameter tuning over predefined search spaces, LLMs can synthesize semantically meaningful algorithmic modifications within bounded code regions~\cite{alphaevolve}. Recent work like OpenEvolve~\cite{openevolve} demonstrates population-based code evolution through iterative LLM-driven mutation and selection, but lacks strategic guidance for complex algorithmic optimization. These observations motivate us to explore LLM-driven code evolution tailored for technology mapping algorithms.

To this end, we propose MappingEvolve, an open-source framework that leverages LLMs to directly evolve the core algorithms of technology mapping. Through systematic analysis of existing mapping implementations, we identify three fundamental operators: MatchPhase (delay and area-flow optimization), MatchPhaseExact (exact-area optimization), and MatchDropPhase (phase unification). They encapsulate critical algorithmic trade-offs while providing well-defined boundaries amenable to controlled evolution. Our framework employs a hierarchical Planner $\rightarrow$ Evolver $\rightarrow$ Evaluator architecture that decouples strategic operator selection from concrete heuristic mutation. To ensure functional correctness, we enforce syntactic and semantic constraints through bounded edit regions and multi-stage validation comprising compilation, logical equivalence checking, and quality-of-result evaluation. A unified performance metric $S_{overall}$ quantifies the area-delay trade-off to guide the evolution process.

Overall, the main contributions of this work are as follows:
\begin{itemize}
  \item To our knowledge, we present the first LLM-driven framework that directly evolves the core algorithmic operators of technology mapping, in contrast to prior approaches limited to external script generation.
  \item We design a hierarchical Planner-Evolver-Evaluator architecture that enforces safety through syntactic boundaries
        and multi-stage validation (compilation, logical equivalence checking, and quality-of-result assessment), enabling safe exploration of the algorithmic design space.
  \item Experimental results demonstrate $11.5\times$ performance improvement over direct operator evolution on ISCAS85 benchmarks \cite{ISCAS85}, achieving 10.04\% area reduction versus ABC~\cite{ABC} and 7.93\% versus mockturtle~\cite{mockturtle}, with 46.6\%--96.0\% $S_{overall}$ improvement on EPFL benchmarks \cite{EPFL2015}.
  \item We release our complete implementation, including source code, evolution prompts, and per-iteration artifacts, to facilitate reproducibility and future research in this field.
\end{itemize}

\section{Technology Mapping Analysis}
\label{sec:preliminaries}

\subsection{Technology Mapping Algorithm}
\label{sec:technology_mapping}
Technology mapping translates a technology-independent Boolean network $G$ with node set $V$, typically an And-Inverter Graph (AIG)~\cite{AIG, ABC}, into a gate-level netlist built from a given standard cell library $L$. Given $G$ and $L$, the goal is to find a mapping $M: V \to L$ that minimizes optimization objectives such as circuit area $A(M)=\sum_{v \in V}\mathrm{area}(M(v))$, worst-case delay $D(M)=\max_{p \in \text{Paths}(G)}$\linebreak$\sum_{v \in p}\mathrm{delay}(M(v))$, or a weighted combination thereof.

\begin{definition}[Logic Phase]
    Each node $v$ can be realized in two output \emph{phases}: positive ($\phi=0$, non-inverted) or negative ($\phi=1$, inverted). When matching a cut $c$ to a gate $g$, the gate's input phases determine the required phase $\phi_l \in \{0,1\}$ for each leaf $l \in c$. If a gate requires a phase not currently implemented by its fanin, an inverter must be inserted.
    \label{def_phase}
\end{definition}

A key challenge in technology mapping is balancing multiple conflicting optimization objectives. Optimizing solely for delay often results in excessive area overhead (due to gate duplication and increased logic depth buffering), while aggressive area minimization can degrade timing performance (by sharing gates across multiple paths, increasing fanout and delay). To address this trade-off, modern technology mappers adopt an iterative refinement strategy that progressively optimizes different objectives across multiple rounds.

To enable LLM-driven evolution, we analyze the mockturtle~\cite{mockturtle} library's implementation and formalize its multi-round mapping process. We identify that the algorithm relies on three optimization metrics applied across different rounds:
\begin{itemize}
    \item \textbf{DelayFlow} records the arrival time at node $v$ for cut $c$ matched to gate $g$:
          \begin{equation}
              T(v, c, g) = \mathrm{delay}(g) + \max_{l \in c} T(l, \phi_l),
              \label{eq_delay_flow}
          \end{equation}
          where $l$ denotes each leaf in cut $c$, $\phi_l \in \{0,1\}$ is the required phase at leaf $l$ (determined by gate $g$'s input phases as defined in \Cref{def_phase}), and $\mathrm{delay}(g)$ denotes the maximum pin-to-output delay across all inputs of gate $g$.

    \item \textbf{AreaFlow} estimates the amortized area contribution:
          \begin{equation}
              F(v, c, g) = \mathrm{area}(g) + \sum_{l \in c} \frac{F(l, \phi_l)}{\mathrm{fo}(l)},
              \label{eq_area_flow}
          \end{equation}
          where $\mathrm{fo}(l)$ is the estimated fanout count of leaf node $l$. This metric recursively accounts for area flow from fanin nodes, assuming each node's cost is shared among its fanouts proportionally.

    \item \textbf{ExactArea} records the true incremental area through reference counting:
          \begin{equation}
              E(v, c, g) = \mathrm{area}(g) + \sum_{\substack{l \in c \\ \rho(l, \phi_l) = 0}} E(l, \phi_l),
              \label{eq_exact_area}
          \end{equation}
          where $\rho(l, \phi_l)$ is the current reference count of leaf $l$ at phase $\phi_l$. Only leaves with zero reference count (i.e., $\rho(l, \phi_l) = 0$) contribute their local area $E(l, \phi_l)$ to the total.
\end{itemize}

This staged optimization, beginning with delay-oriented optimization using $\mathrm{DelayFlow}(T)$, transitioning to area flow optimization using $\mathrm{AreaFlow}(F)$, and culminating in exact area optimization using $\mathrm{ExactArea}(E)$, allows the algorithm to efficiently explore quality-runtime trade-offs.

\begin{algorithm}[t]
    \small
    \setlength{\hsize}{0.95\linewidth}
    \KwIn{Logic network $G$, standard cell library $L$}
    \KwOut{Mapped netlist $N$}
    \KwData{$R_{delay}$: delay rounds, $R_{flow}$: area flow rounds, $R_{exact}$: exact area rounds, $R = R_{delay} + R_{flow} + R_{exact}$}
    \tcp{Iterative optimization with varying objectives}
    \For{\textup{round $r = 1$ to $R$}}{
        \For{\textup{each node $v \in G$ in topological order}}{
            \eIf{$r \le R_{delay} + R_{flow}$}{
                \tcp{Delay and area flow optimization rounds}
                $\mathrm{MatchPhase}(v, \text{phase}=0, L)$\;
                $\mathrm{MatchPhase}(v, \text{phase}=1, L)$\;
            } {
                \tcp{Exact area optimization rounds}
                $\mathrm{MatchPhaseExact}(v, \text{phase}=0, L)$\;
                $\mathrm{MatchPhaseExact}(v, \text{phase}=1, L)$\;
            }
            $\mathrm{MatchDropPhase}(v)$\;
        }
        Update mapping coverage and compute metrics\;
        Backward propagate required times\;
    }
    Finalize netlist $N$ from selected cuts\;
    \Return{$N$}
    \caption{Technology mapping with three key operators.}
    \label{alg:tech_mapping}
\end{algorithm}

\begin{figure*}[t]
    \centering
    \includegraphics[width=0.98\linewidth]{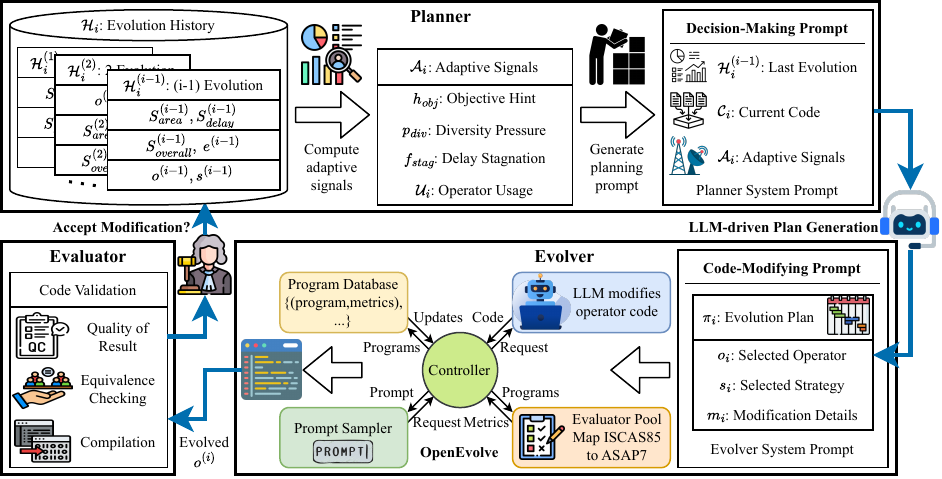}
    \caption{The overall flow of MappingEvolve.}
    \label{fig_3_flow}
\end{figure*}

\subsection{Operator Abstraction}
\label{sec:operator_abstraction}
To enable LLM-driven evolution of technology mapping algorithms, we systematically analyze the implementation and abstract the iterative optimization process into a unified algorithmic framework shown in \Cref{alg:tech_mapping}. The algorithm executes $R$ optimization rounds, where $R = R_{delay} + R_{flow} + R_{exact}$ combines delay rounds ($R_{delay}$, typically 1), area flow rounds ($R_{flow}$, typically 1-2), and exact area rounds ($R_{exact}$, typically 2-3). Each round processes all nodes in topological order.
Critically, we identify that the mapping quality is determined by three core operators repeatedly invoked across all rounds:
\begin{itemize}
    \item \textbf{MatchPhase} (Lines 4-5): For a given node $v$ and output phase $\phi \in \{0, 1\}$, this operator evaluates all feasible cuts $\mathcal{C}(v)$ and their compatible library gates. For each cut $c \in \mathcal{C}(v)$ and gate $g \in L$, it computes the arrival time $T(v, c, g)$ (via \Cref{eq_delay_flow}) and area flow $F(v, c, g)$ (via \Cref{eq_area_flow}), then selects the best match by minimizing a cost function $C(c, g) = \alpha_r \cdot T(v, c, g) + (1-\alpha_r) \cdot F(v, c, g)$, where $\alpha_r \in [0,1]$ determines the delay-area trade-off priority in round $r$. In practice, $\alpha_r$ starts close to 1 in delay-oriented rounds (emphasizing $T$) and decreases toward 0 in area-flow rounds (increasing weight on $F$). This operator is invoked in delay and area flow optimization rounds (when $r \leq R_{delay} + R_{flow}$).

    \item \textbf{MatchPhaseExact} (Lines 7-8): This operator serves the same purpose as \text{MatchPhase} but performs exact area computation by recursively dereferencing the previous best cut from the current mapping cover and then evaluating each candidate cut through recursive reference counting (via \Cref{eq_exact_area}). Unlike the flow-based heuristics in \text{MatchPhase}, this yields the true incremental area $E(v,c,g)$ but at a higher computational cost. In exact-area rounds, candidate selection minimizes $E(v,c,g)$ directly under timing feasibility constraints. It replaces \text{MatchPhase} in the exact area optimization rounds (when $r > R_{delay} + R_{flow}$).

    \item \textbf{MatchDropPhase} (Line 9): After both output phases have been matched, this operator attempts to unify them by checking whether a single gate match combined with an output inverter can cover both phases at a lower total cost. If unification reduces cost without violating timing constraints, it eliminates the redundant phase match. This operator is applied in every optimization round.
\end{itemize}

After each round, the algorithm updates the mapping coverage and computes required arrival times to guide subsequent rounds (Lines 10-11).

\subsection{LLM-driven Evolution Target Selection}
\label{sec:target_selection}

The three operators identified above, $\mathcal{O}=$\{\text{MatchPhase}, \text{MatchPhaseExact}, \text{MatchDropPhase}\}, encapsulate the core optimization logic of technology mapping algorithms. We select these operators as ideal targets for LLM-driven code evolution because they exhibit three key properties that balance evolvability with safety:

First, each operator implements \textbf{self-contained optimization heuristics} with tunable parameters and trade-off decisions (e.g., cost function weights in \text{MatchPhase}, timing slack thresholds in \text{MatchPhaseExact}, phase unification criteria in \text{MatchDropPhase}), precisely the type of algorithmic choices where LLMs explore strategies beyond human-designed rules.

Second, the mockturtle library~\cite{mockturtle} provides \textbf{well-defined modular interfaces}, allowing us to isolate each operator in a separate C++ source file. We designate safe modification regions using \texttt{EVOLVE-BLOCK} markers, ensuring LLM edits are confined to optimization logic (e.g., cost function computation) while preserving API integrity (e.g., class member access).

Third, operator modifications directly impact \textbf{measurable performance metrics} (area $A(M)$ and delay $D(M)$), enabling quantitative feedback for iterative evolution. Combined with compilation and equivalence checks, this establishes a robust validation process.

By abstracting the mapping algorithm into these three evolvable operators with bounded modification regions, we create a structured search space that balances exploration freedom with safety guarantees, a key enabler for systematic LLM-driven algorithm improvement.

\section{MappingEvolve}
\label{sec:method}
This section presents MappingEvolve, a framework that leverages LLMs to evolve technology mapping algorithms, as shown in \Cref{fig_3_flow}.
Based on the operator abstraction established in \Cref{sec:operator_abstraction}, MappingEvolve employs a hierarchical Planner$\to$Evolver$\to$Evaluator architecture to iteratively evolve the three core operators. Unlike direct code generation approaches that lack strategic guidance, this architecture decouples strategic planning from implementation, enabling targeted exploration of algorithmic improvements while maintaining code safety through bounded modification regions and multi-stage validation.

\subsection{Optimization Objectives}
\label{sec:objectives}

As formalized in \Cref{sec:technology_mapping}, technology mapping seeks to minimize circuit area $A(M)$ and delay $D(M)$ for a given mapping $M$. These objectives often conflict: aggressive area minimization may increase delay, while delay-optimal mappings frequently require extra gates.

To guide the evolution process and quantify the area-delay trade-off, we define four metrics for evaluating evolved mapping $M'$ relative to the original $M$:
\begin{itemize}
    \item $S_{area}(M')=\frac{A(M) - A(M')}{A(M)}$: Normalized area reduction.
    \item $S_{delay}(M')=\frac{D(M) - D(M')}{D(M)}$: Normalized delay reduction.
    \item $S_{overall}(M') = \alpha \cdot S_{area}(M') + (1 - \alpha) \cdot S_{delay}(M')$: Weighted combination where $\alpha \in [0, 1]$ controls area-delay priority.
    \item $e(M') \in [0,1]$: Logical equivalence failure rate (must be zero for valid implementations).
\end{itemize}

To navigate the area-delay trade-off, we define three optimization strategies $\mathcal{S}$:
\begin{itemize}
    \item \textbf{Area-opt} prioritizes area reduction while maintaining timing feasibility.
    \item \textbf{Delay-opt} prioritizes delay reduction, tolerating modest area increases.
    \item \textbf{Balanced} seeks simultaneous improvements via $S_{overall}$.
\end{itemize}
Strategy selection adapts to performance signals from recent iterations: when area improves but delay consistently degrades, \textbf{delay-opt} restores timing quality; when area shows severe degradation, \textbf{area-opt} focuses on area optimization; when both objectives show no clear trend, \textbf{balanced} explores the Pareto frontier.

\subsection{Hierarchical Agent Architecture}

The framework comprises three components: a \textbf{Planner} selecting optimization targets based on evolution history, an \textbf{Evolver} generating plan-conditioned mutations, and an \textbf{Evaluator} validating implementations. This decouples strategic planning from code generation and quality assessment.

\subsubsection{Planner}
The Planner prepares structured inputs for an LLM-based decision maker and specifies the required output format for downstream Evolver. It decouples \emph{what to optimize} (strategic planning) from \emph{how to implement} (code generation), enabling systematic exploration while avoiding local optima.

At iteration $i$, the Planner processes evolution history $\mathcal{H}_i = \{(S_{area}^{(j)}, S_{delay}^{(j)}, S_{overall}^{(j)}, e^{(j)}, o^{(j)}, s^{(j)})\}_{j=\max(1,i-w)}^{i-1}$ of the most recent $w$ iterations, where $o^{(j)} \in \mathcal{O}$ is the modified operator at iteration $j$ and $s^{(j)} \in \mathcal{S}$ is the employed strategy. It extracts current code context $\mathcal{C}_i = \{c_o\}_{o \in \mathcal{O}}$ (where $c_o$ is operator $o$'s implementation) and computes adaptive signals $\mathcal{A}_i=\{h_{obj}, p_{div}, f_{stag}, \mathcal{U}_i\}$ from $\mathcal{H}_i$ to guide decision-making:
\begin{itemize}
    \item \emph{Objective hint} $h_{obj} \in \{\text{delay}, \text{area}, \text{balanced}\}$ guides strategy prioritization, computed as:
          \begin{align}
              h_{obj}             & = \begin{cases}
                                          \text{area}     & \text{if } (n_{area} > 0 \land \bar{\Delta}_{delay} \geq 0)      \\
                                          \text{delay}    & \text{else if } (n_{delay} > 0 \land \bar{\Delta}_{area} \geq 0) \\
                                          \text{balanced} & \text{otherwise}
                                      \end{cases},                                                \\
              n_{area}            & = |\{j \in \mathcal{H}_i : S_{area}^{(j)} > S_{area}^{(j-1)}\}|, n_{delay} \text{ (similarly)},                                     \\
              \bar{\Delta}_{area} & = \frac{1}{|\mathcal{H}_i|}\sum_{j \in \mathcal{H}_i}(S_{area}^{(j)} - S_{area}^{(j-1)}), \bar{\Delta}_{delay} \text{ (similarly)}.
          \end{align}
          If one metric improves while the other does not worsen, the hint guides continued optimization of the improving metric (e.g., if area reduces but delay remains stable, $h_{obj} = \text{area}$).
    \item \emph{Diversity pressure} $p_{div} = \begin{cases}
                  \text{HIGH} & \text{if } n_{cs} \geq \lceil w/2 \rceil \\
                  \text{LOW}  & \text{otherwise}
              \end{cases}$ detects operator over-exploitation,
          where $n_{cs}$ counts consecutive selections of the same operator. If $p_{div} = \text{HIGH}$, least-used operator $o_i \in \{o \in \mathcal{O}: n_o = \min_{o' \in \mathcal{O}} n_{o'}\}$ is selected to escape local optima by exploring complementary operators rather than over-exploiting a single search space.
    \item \emph{Delay stagnation} $f_{stag} = \mathbb{I}[n_{area} > 0 \land \forall j \in \mathcal{H}_i, S_{delay}^{(j)} < 0]$ flags dangerous trade-off patterns when area gains consistently sacrifice timing. If $f_{stag} = 1$, $s_i = \text{delay-opt}$.
    \item \emph{Operator usage} $\mathcal{U}_i = \{(o, n_o)\}_{o \in \mathcal{O}}$, where $n_o = |\{j \in \mathcal{H}_i : o^{(j)} = o\}|$, tracks selection frequency to identify under-explored operators.
\end{itemize}

The Planner constructs a decision-making prompt ($\mathcal{H}_i^{(i-1)}$, $\mathcal{C}_i$, $\mathcal{A}_i$) and sends it to an LLM, where $\mathcal{H}_i^{(i-1)}$ contains performance data from iteration $i-1$. The Planner specifies the required output format: evolution plan $\pi_i = (o_i, s_i, m_i)$, where $o_i \in \mathcal{O}$ is the selected operator, $s_i \in \mathcal{S}$ is the optimization strategy, and $m_i$ describes the modification (target code region, implementation approach, expected impact $\Delta_{exp} \in \mathbb{R}^2$, and constraints). The LLM generates $\pi_i$ by reasoning over the previous iteration's results, code structure, and adaptive signals. The Planner validates the output format and forwards $\pi_i$ to the Evolver for implementation.

\subsubsection{Evolver}
The Evolver translates the Planner's evolution plan $\pi_i = (o_i, s_i, m_i)$ into concrete code modifications through plan-conditioned population evolution. Unlike general-purpose code generation that explores the full modification space $\mathcal{X}_{all}$, the Evolver constrains search to a strategically reduced subspace $\mathcal{X}_{o_i, s_i} \subset \mathcal{X}_{all}$ aligned with $\pi_i$.

\minisection{Plan-Guided Search Space Reduction}
We implement the Evolver using OpenEvolve~\cite{openevolve}, a population-based LLM code evolution framework that maintains multiple candidate solutions and employs island-based parallel evolution with iterative mutation and selection. Crucially, we extend OpenEvolve from general-purpose code generation to plan-conditioned evolution by injecting $\pi_i$ into its system prompt as structured constraints:
\begin{itemize}
    \item \textbf{Target Region}: $m_i$ specifies the code region in current implementation $c_{o_i}\in\mathcal{C}_i$ of the selected operator $o_i$ to confine mutations.
    \item \textbf{Implementation Guidance}: $m_i$ provides algorithmic hints aligned with $s_i$ (e.g., for \textbf{delay-opt}, ``prioritize cuts with minimal depth increase'').
    \item \textbf{Expected Impact}: $\Delta_{exp} = (\Delta_{area}, \Delta_{delay})$ filters candidate solutions during evaluation.
    \item \textbf{Semantic Constraints}: $m_i$ encodes operator invariants (e.g., ``ensure matching respects library gate fanin limits'') to prevent invalid mutations.
\end{itemize}
This transforms OpenEvolve from undirected search into a strategic exploration engine: the population evolves within $\mathcal{X}_{o_i, s_i}$ defined by both \emph{where} (operator $o_i$, code region) and \emph{how} (strategy $s_i$, implementation hints), improving convergence while maintaining diversity through island-based parallelism.

\minisection{Multi-Layered Safety Enforcement}
The Evolver enforces safety through hierarchical constraints:
\begin{enumerate}
    \item \textbf{Syntactic Boundaries}: Edits confined to \texttt{EVOLVE-BLOCK} regions containing only optimization heuristics, excluding function signatures, API interfaces, and data structures.
    \item \textbf{Semantic Constraints}: $m_i$ specifies algorithmic invariants (e.g., ``$k$-LUT input limit'') validated before compilation, filtering invalid mutations early.
    \item \textbf{Incremental Validation}: OpenEvolve's island evolution creates a feedback loop (compilation $\to$ equivalence $\to$ QoR) through the Evaluator, enabling convergence to valid solutions.
\end{enumerate}
This combination of plan-guided search reduction and multi-layered constraints enables complex modifications while maintaining correctness. It is critical for technology mapping, as subtle logic errors produce functionally incorrect circuits undetected by compilation.

\subsubsection{Evaluator}
\label{sec:evaluator}
The Evaluator receives evolved operator code from the Evolver, merges it with the unchanged operators from $\mathcal{C}_i$ to form a complete mapper implementation, then validates the integrated code through a three-stage pipeline and generates a reward signal $R$ that quantifies modification quality. Each candidate undergoes hierarchical validation with corresponding penalties:
\begin{equation}
    R = \begin{cases}
        r_{compile}                         & \text{if compilation fails}         \\
        r_{equiv} - \beta \cdot e           & \text{if equivalence fails, } e > 0 \\
        \max(r_{equiv}, S_{overall})        & \text{if } S_{overall} < 0          \\
        \frac{S_{overall}}{1 + S_{overall}} & \text{if } S_{overall} \geq 0
    \end{cases},
\end{equation}
where $r_{equiv} < 0$, $\beta > 0$, and $r_{compile} = r_{equiv} - \beta $. This hierarchical design maps validation stages to differentiated feedback:
\begin{itemize}
    \item \textbf{Compilation} ($r_{compile}$): Merges the evolved operator $c_{o_i}$ with unchanged operators from $\mathcal{C}_i$ into a complete mapper, then builds the integrated code and returns diagnostics. Failures receive the harshest penalty, immediately rejecting syntactically invalid code.
    \item \textbf{Equivalence Checking} ($r_{equiv} - \beta \cdot e$): Verifies output netlist functional equivalence using ABC~\cite{ABC}'s combinational equivalence checker \texttt{cec}. This is critical for technology mapping, where performance optimizations are meaningless if circuit functionality is corrupted. Violations incur strong penalties proportional to the failure rate $e$, discouraging semantic errors while allowing partial progress.
    \item \textbf{Quality of Result} ($\max(r_{equiv}, S_{overall})$ or $\frac{S_{overall}}{1 + S_{overall}}$): Evaluates the validated mapper $M'$ on benchmark suite $\mathcal{B}$ and computes $S_{area}$, $S_{delay}$, $S_{overall}$ as defined in \Cref{sec:objectives}. For regressions ($S_{overall} < 0$), capping at $r_{equiv}$ prevents catastrophic penalties for valid but suboptimal code, enabling exploration of temporary regressions. For improvements ($S_{overall} \geq 0$), sigmoid transformation bounds rewards to $(0, 1)$ to prevent unbounded optimization bias.
\end{itemize}

The evaluation record $(S_{area}^{(i)}, S_{delay}^{(i)}, S_{overall}^{(i)}, e^{(i)}, o^{(i)}, s^{(i)})$ enters the evolution history $\mathcal{H}_{i+1}$. The Planner analyzes this accumulated history through adaptive signals $\mathcal{A}_{i+1}$, establishing a strategic feedback loop where Evaluator results at iteration $i$ shape the Planner's inputs for iteration $i+1$.

\subsection{Iterative Evolution Loop}

The evolution process operates in iterations. Each iteration executes the following steps:
\begin{enumerate}
    \item The Planner analyzes performance history and proposes an evolution plan.
    \item The Evolver implements the proposed modifications to the selected operator.
    \item The Evaluator merges modified code without any changes, then compiles, tests, and measures the evolved mapper.
    \item MappingEvolve applies a multi-criteria acceptance policy to decide whether to accept the modification, then updates the state and feeds results back to the Planner. A modification is accepted if it achieves sufficient reward ($R_i \geq \tau$) or demonstrates significant delay improvement ($S_{delay}^{(i)} \geq \gamma \cdot S_{delay}^{best}$, where $\gamma \in (0, 1]$). The latter criterion preserves delay optimizations even when the overall reward is suboptimal, recognizing that delay reduction is harder to achieve than area optimization. State updates include:
          \begin{itemize}
              \item \textbf{Code state} $\mathcal{C}_{i+1}$: If accepted, $c_{o_i}^{(i+1)} \leftarrow$ evolved code; otherwise $c_{o_i}^{(i+1)} \leftarrow c_{o_i}^{(i)}$. The code for unselected operators remains unchanged.
              \item \textbf{Evolution history} $\mathcal{H}_{i+1}$: Appends the current evaluation record ($S_{area}^{(i)}$, $S_{delay}^{(i)}$, $S_{overall}^{(i)}$, $e^{(i)}$, $o^{(i)}$, $s^{(i)}$), retaining only the most recent $w$ iterations.
              \item \textbf{Best performance} $S_{delay}^{best}$: Updates to $\max(S_{delay}^{best}, S_{delay}^{(i)})$ for acceptance criteria.
          \end{itemize}
\end{enumerate}

The evolution continues for a maximum of $N$ iterations or no improvement over consecutive iterations.

\section{Experimental Results}
\label{sec:experiments}

\begin{table}[t]
    \centering
    \small
    \caption{Comparison of $S_{overall}$ and $e$ scores across different models and methods on ISCAS85 benchmarks \cite{ISCAS85}.}
    \label{tab_comparison_LLMs}
    \begin{tabular}{lc|c|c|c}
        \toprule
        \multicolumn{2}{c|}{Model}                          & Base Model      & $S_{overall}$                & $e$         \\
        \midrule
        \multirow{3}{*}{OpenEvolve}                         & MatchPhase      & \multirow{3}{*}{DeepSeek-V3} & 0.00 & 0.00 \\
                                                            & MatchPhaseExact &                              & 0.00 & 0.00 \\
                                                            & MatchDropPhase  &                              & 0.02 & 0.09 \\
        \midrule
        \multicolumn{2}{c|}{\multirow{3}{*}{MappingEvolve}} & DeepSeek-V3     & 0.23                         & 0.00        \\
                                                            &                 & Qwen3-Max                    & 0.18 & 0.00 \\
                                                            &                 & GPT-5                        & 0.30 & 0.00 \\
        \bottomrule
    \end{tabular}
\end{table}

\begin{table*}[t]
    \centering
    \footnotesize
    \caption{Technology mapping results on EPFL benchmarks \cite{EPFL2015} with ASAP7 standard cell library \cite{ASAP7}.\vspace{4pt}}
    \label{tab_mapping_epfl}
    \begin{tabular}{c|rr|rrr|rrr|rrr|rrr}
        \toprule
        \multirow{3}{*}{Circuit} & \multicolumn{2}{c|}{\multirow{2}{*}{AIG}}   & \multicolumn{3}{c|}{\multirow{2}{*}{ABC}} & \multicolumn{3}{c|}{\multirow{2}{*}{mockturtle}} & \multicolumn{6}{c}{MappingEvolve (Ours)}                                                                                                                                     \\ \cline{10-15}
                                 & \multicolumn{2}{c|}{}                       & \multicolumn{3}{c|}{}                     & \multicolumn{3}{c|}{}                            & \multicolumn{3}{c|}{DeepSeek-V3}         & \multicolumn{3}{c}{GPT-5}                                                                                                         \\
                                 & size                                        & depth                                     & area                                             & delay                                    & t(s)                                & area     & delay     & t(s)   & area     & delay     & t(s)   & area     & delay     & t(s) \\
        \midrule
        adder                    & 1019                                        & 255                                       & 100.53                                           & 2574.36                                  & 0.05                                & 92.42    & 2574.36   & 0.06   & 76.94    & 2583.24   & 0.02   & 90.6     & 2768.87   & 0.06 \\
        bar                      & 3141                                        & 12                                        & 263.63                                           & 169.9                                    & 0.08                                & 298.82   & 171.90    & 0.07   & 297.54   & 171.9     & 0.08   & 225.74   & 180.16    & 0.07 \\
        div                      & 40633                                       & 4394                                      & 4021.98                                          & 43768.73                                 & 0.76                                & 3510.25  & 43295.70  & 1.11   & 2930.53  & 45227.7   & 1.26   & 3012.57  & 43543.04  & 1.14 \\
        hyp                      & 211329                                      & 24893                                     & 16756.18                                         & 198844.69                                & 5.4                                 & 15895.39 & 197081.23 & 7.05   & 13996.27 & 238083.28 & 7.27   & 15296.85 & 198729.45 & 6.51 \\
        log2                     & 29371                                       & 387                                       & 2019.22                                          & 3990.67                                  & 1                                   & 2047.72  & 3968.44   & 2.11   & 1641.84  & 4642.54   & 2.59   & 1853.82  & 4126.81   & 2.4  \\
        max                      & 2832                                        & 206                                       & 256.14                                           & 2101.85                                  & 0.1                                 & 225.89   & 2101.13   & 0.12   & 183.96   & 2151.64   & 0.19   & 202.65   & 2220.27   & 0.18 \\
        multiplier               & 24556                                       & 262                                       & 1816.61                                          & 2730.91                                  & 0.54                                & 1913.22  & 2662.96   & 0.82   & 1632.12  & 2956.62   & 1      & 1743.2   & 2673.95   & 0.88 \\
        sin                      & 5041                                        & 179                                       & 433.57                                           & 1843.39                                  & 0.18                                & 405.80   & 1832.67   & 0.37   & 324.23   & 2032.53   & 0.47   & 367.25   & 1910.26   & 0.4  \\
        sqrt                     & 18368                                       & 6048                                      & 1525.72                                          & 49967.46                                 & 0.67                                & 1484.04  & 47964.09  & 0.56   & 1206.42  & 57824.42  & 0.7    & 1586.17  & 55116.87  & 0.59 \\
        square                   & 16623                                       & 248                                       & 1243                                             & 2509.4                                   & 0.43                                & 1208.58  & 2509.22   & 0.39   & 1161.13  & 2856.92   & 0.45   & 1149.15  & 2519.15   & 0.41 \\
        arbiter                  & 11839                                       & 87                                        & 783.69                                           & 898.75                                   & 0.16                                & 766.44   & 898.75    & 1.65   & 766.44   & 898.75    & 1.78   & 591.09   & 969.11    & 1.67 \\
        cavlc                    & 662                                         & 16                                        & 41.19                                            & 188.71                                   & 0.05                                & 40.69    & 185.77    & 0.03   & 40.86    & 185.77    & 0.03   & 38.34    & 191.66    & 0.03 \\
        ctrl                     & 108                                         & 8                                         & 7.51                                             & 104.36                                   & 0.06                                & 6.92     & 103.17    & 0.00   & 6.9      & 103.17    & 0.01   & 6.9      & 106.33    & 0.01 \\
        dec                      & 304                                         & 3                                         & 28.73                                            & 66.15                                    & 0.06                                & 30.83    & 65.72     & 0.06   & 29.75    & 65.72     & 0.08   & 27.59    & 66.15     & 0.07 \\
        i2c                      & 1161                                        & 15                                        & 70.43                                            & 165.99                                   & 0.07                                & 70.23    & 164.10    & 0.06   & 70.23    & 164.1     & 0.07   & 69.13    & 174.18    & 0.07 \\
        int2float                & 214                                         & 15                                        & 13.39                                            & 181.17                                   & 0.05                                & 12.80    & 181.00    & 0.01   & 12.8     & 181       & 0.02   & 12.3     & 193.8     & 0.01 \\
        mem\_ctrl                & 45547                                       & 106                                       & 2798.58                                          & 1015.48                                  & 0.89                                & 2716.20  & 1006.24   & 1.84   & 2623.27  & 1072.94   & 2.15   & 2631.09  & 1040.39   & 1.85 \\
        priority                 & 683                                         & 214                                       & 52.63                                            & 2155.94                                  & 0.07                                & 52.20    & 2155.87   & 0.03   & 57.73    & 2288.23   & 0.03   & 50.81    & 2264.96   & 0.04 \\
        router                   & 182                                         & 18                                        & 12.7                                             & 187.2                                    & 0.06                                & 12.85    & 187.20    & 0.02   & 12.93    & 187.2     & 0.03   & 12.25    & 192.22    & 0.03 \\
        voter                    & 9654                                        & 59                                        & 1004.29                                          & 729.37                                   & 0.22                                & 954.64   & 723.58    & 0.37   & 705.63   & 843.44    & 0.37   & 761.58   & 773.44    & 0.38 \\ \midrule
        Ave. ratio               & \multicolumn{2}{c|}{\multirow{2.5}{*}{---}} & 1                                         & 1                                                & 1                                        & 0.9771                              & 0.9926   & 1.5474    & 0.8996 & 1.0571   & 1.7550    & 0.8996 & 1.0369   & 1.6557           \\
        \cline{1-1}\cline{4-15}
        $S_{overall}$            & \multicolumn{2}{c|}{}                       & \multicolumn{3}{c|}{0}                    & \multicolumn{3}{c|}{0.0174}                      & \multicolumn{3}{c|}{0.0255}              & \multicolumn{3}{c}{\textbf{0.0341}}                                                                                               \\ \bottomrule
    \end{tabular}
\end{table*}

\subsection{Experimental Setup}

\subsubsection{Implementation Details}
MappingEvolve is implemented in Python and evaluated on Ubuntu 22.04 with an Intel Xeon Gold 6226R CPU @ 2.90GHz and 256GB memory. We use OpenEvolve~\cite{openevolve} as the Evolver component, executing $N=30$ outer iterations with 3 OpenEvolve iterations per step. During evolution, candidates are validated on ISCAS85 benchmarks~\cite{ISCAS85} (11 circuits), where failure rate $e$ measures the ratio of non-equivalent circuits. For QoR evaluation, circuits are optimized via ABC's \texttt{compress2} and then mapped to ASAP7~\cite{ASAP7} using the evolved mapper. We evaluate using DeepSeek-V3~\cite{deepseekv3}, Qwen3-Max~\cite{qwen3}, and GPT-5~\cite{gpt5} as base LLMs.

\minisection{Hyperparameters} $\alpha = 0.5$ (balanced area-delay weighting), $\tau = -0.1$ (reward threshold allowing slight degradation), $\gamma = 0.8$ (delay preservation factor), $w=5$ (history window). Reward function: $r_{equiv} = -0.4$, $\beta = 0.1$, $r_{compile} = -0.5$, thus $R \in [-0.5, 0.5]$.

\minisection{Computational Cost} The entire evolution process of MappingEvolve takes \textbf{$\sim 1$ million tokens ($\mathdollar 10$)} and \textbf{$\sim 1.5$ hours}.

\subsubsection{Benchmarks and Baselines}
We use ISCAS85 benchmarks~\cite{ISCAS85} for ablation studies (\Cref{tab_comparison_LLMs}) and EPFL benchmarks~\cite{EPFL2015} for performance comparison (\Cref{tab_mapping_epfl}).

\minisection{Experiment 1: Framework Effectiveness} We compare MappingEvolve against OpenEvolve using DeepSeek-V3~\cite{deepseekv3}. OpenEvolve evolves each operator (MatchPhase, MatchPhaseExact, MatchDropPhase) separately for 90 iterations (matching $30 \times 3 = 90$ total OpenEvolve calls in MappingEvolve for a fair comparison).

\minisection{Experiment 2: Model Generalization} We evaluate MappingEvolve with different base LLMs, including DeepSeek-V3~\cite{deepseekv3}, Qwen3-Max~\cite{qwen3}, and GPT-5~\cite{gpt5}.

\minisection{Experiment 3: Performance Comparison} We assess the evolved mappers against ABC (\texttt{\&nf})~\cite{ABC} and mockturtle (\texttt{map})~\cite{mockturtle} on EPFL benchmarks~\cite{EPFL2015}. All netlists pass ABC's \texttt{cec} equivalence checking, ensuring functional correctness.

\subsection{Results and Analysis}

\subsubsection*{Effectiveness of Hierarchical Architecture}
\Cref{tab_comparison_LLMs} compares OpenEvolve and MappingEvolve on ISCAS85 using DeepSeek-V3. OpenEvolve achieves minimal improvements: MatchPhase and MatchPhaseExact yield $S_{overall}=0.00$, while MatchDropPhase reaches only $S_{overall}=0.02$ but introduces equivalence failures ($e=0.09$). In contrast, MappingEvolve achieves $S_{overall}=0.23$ with $e=0.00$, an $11.5\times$ improvement while maintaining perfect correctness. This validates that our architecture effectively coordinates cross-operator evolution through strategic planning and adaptive selection.

\subsubsection*{Model Generalization Across Different LLMs}
\Cref{tab_comparison_LLMs} evaluates MappingEvolve with three LLMs on ISCAS85. GPT-5 achieves $S_{overall}=0.30$, followed by DeepSeek-V3 ($0.23$) and Qwen3-Max ($0.18$), all with $e=0.00$. Even Qwen3-Max outperforms the OpenEvolve baseline ($0.02$) by $9\times$, confirming framework generalization across model families.

\subsubsection*{Comparison with State-of-the-Art Tools on EPFL Benchmarks}
\Cref{tab_mapping_epfl} compares evolved mappers against ABC (\texttt{\&nf})~\cite{ABC} and mockturtle (\texttt{map})~\cite{mockturtle} on EPFL benchmarks~\cite{EPFL2015}.
MappingEvolve achieves \textbf{10.04\% area reduction} versus ABC and \textbf{7.93\% area reduction} versus mockturtle, with a trade-off of 4.46\%-6.50\% delay increase. The $S_{overall}$ metric shows 46.6\%-96.0\% improvement over mockturtle,
demonstrating that despite the delay trade-off, the evolved mappers discover beneficial area-delay trade-offs. This reflects our balanced area-delay weighting ($\alpha=0.5$), enabling MappingEvolve to effectively navigate the area-delay Pareto frontier.

\subsubsection*{Analysis of Evolved Code Improvements}
To understand how MappingEvolve discovers effective optimizations, we analyze the GPT-5 evolved mapper at iteration 29 ($S_{overall}=0.298$), which demonstrates coordinated improvements across three operators:
\begin{itemize}[leftmargin=*,itemsep=2pt,topsep=3pt]
    \item \textbf{MatchPhase (Delay round):} Introduces area tolerance $t_{tol}=0.25\cdot\mathrm{area}(\text{inv})$, accepting delay-improving cuts only if: (1) area increase $\leq t_{tol}$, OR (2) delay gain $\geq 0.5\cdot\mathrm{delay}(\text{inv})$.
    \item \textbf{MatchPhase (Area round):} Adds area slack $t_{slack}=0.5\cdot\mathrm{area}(\text{inv})$ for delay-improving cuts, enabling opportunistic timing recovery.
    \item \textbf{MatchPhaseExact:} Applies area gain threshold $t_{gain}=0.5\cdot\mathrm{area}(\text{inv})$, accepting area-improving cuts that worsen delay only if area gain $\geq t_{gain}$.
    \item \textbf{MatchDropPhase (Delay round):} Enforces zero area tolerance for phase consolidation, strictly preserving area during delay optimization.
\end{itemize}

\section{Conclusion}
\label{sec:conclusion}

We presented MappingEvolve, a framework that leverages LLMs to evolve technology mapping algorithms through hierarchical planning, controlled mutation, and rigorous validation. By abstracting the mapping process into three evolvable operators, we enable systematic algorithmic exploration while maintaining functional correctness. Experimental results demonstrate the framework's effectiveness: MappingEvolve achieves $11.5\times$ improvement over direct operator evolution on ISCAS85 benchmarks with perfect logical equivalence. On EPFL benchmarks, the evolved mapper achieves 10.04\% area reduction versus ABC, and $S_{overall}$ improvements of 46.6\%--96.0\% over mockturtle, effectively navigating the area-delay trade-off. We release all code, prompts, and evolution artifacts to facilitate future research. Promising directions include delay-oriented optimization and extending to other EDA tools. 

\section*{Acknowledgments}
The research work described in this paper was conducted in the JC STEM Lab of Intelligent Design Automation funded by The Hong Kong Jockey Club Charities Trust and was supported in part by the Research Grants Council of Hong Kong SAR (Grant No.~CUHK14207523).

\bibliographystyle{IEEEtran}
\bibliography{Top,reference}

@string{aspdac   = "IEEE/ACM Asia and South Pacific Design Automation Conference (ASPDAC)"}

@string{dac      = "ACM/IEEE Design Automation Conference (DAC)"}

@string{date     = "IEEE/ACM Proceedings Design, Automation and Test in Eurpoe (DATE)"}

@string{fpga     = "ACM International Symposium on Field-Programmable Gate Arrays (FPGA)"}

@string{iccad    = "IEEE/ACM International Conference on Computer-Aided Design (ICCAD)"}

@string{iwls     = "IEEE/ACM International Workshop on Logic Synthesis"}

@string{mdat     = "IEEE Design \& Test"}

@string{tcad     = "IEEE Transactions on Computer-Aided Design of Integrated Circuits and Systems (TCAD)"}

@string{todaes   = "ACM Transactions on Design Automation of Electronic Systems (TODAES)"}

@string{arxiv    = "arXiv preprint"}

@article{ISCAS85,
  author    = {Hansen, Mark C. and Yalcin, Hakan and Hayes, John P.},
  title     = {Unveiling the {ISCAS-85} Benchmarks: A Case Study in Reverse Engineering},
  year      = {1999},
  publisher = {IEEE Computer Society Press},
  volume    = {16},
  number    = {3},
  issn      = {0740-7475},
  doi       = {10.1109/54.785838},
  journal   = mdat,
  pages     = {72--80},
  numpages  = {9}
}

@inproceedings{EPFL2015,
  title     = {The {EPFL} Combinational Benchmark Suite},
  author    = {Amarù, Luca and Gaillardon, Pierre-Emmanuel and De  Micheli, Giovanni},
  booktitle = iwls,
  year      = {2015}
}

@inproceedings{ASAP7,
  author    = {Vashishtha, Vinay and Vangala, Manoj and Clark, Lawrence T.},
  booktitle = iccad,
  title     = {{ASAP7} predictive design kit development and cell design technology co-optimization: Invited paper},
  year      = {2017},
  volume    = {},
  number    = {},
  pages     = {992-998},
  doi       = {10.1109/ICCAD.2017.8203889}
}

@inproceedings{alan2007priority,
  title     = {Combinational and Sequential Mapping with Priority Cuts},
  booktitle = iccad,
  author    = {Mishchenko, Alan and {Sungmin Cho} and {Satrajit Chatterjee} and Brayton, Robert},
  year      = 2007,
  pages     = {354--361},
  publisher = {IEEE},
  issn      = {1092-3152},
  doi       = {10.1109/ICCAD.2007.4397290},
  isbn      = {978-1-4244-1381-2 978-1-4244-1382-9}
}

@inproceedings{neto2021slap,
  title     = {{SLAP}: A Supervised Learning Approach for Priority Cuts Technology Mapping},
  booktitle = dac,
  author    = {Neto, Walter Lau and Moreira, Matheus T. and Li, Yingjie and Amaru, Luca and Yu, Cunxi and Gaillardon, Pierre Emmanuel},
  year      = 2021,
  volume    = {2021-December},
  pages     = {859--864},
  doi       = {10.1109/DAC18074.2021.9586230},
  isbn      = {978-1-6654-3274-0}
}

@inproceedings{chigarapally2025LEAP,
  author    = {Chigarapally, Chandrabhusan Reddy and Bhakkad, Harshwardhan Nitin and Chowdhury, Animesh Basak and Karfa, Chandan and Bhattacharjee, Sukanta},
  title     = {{LEAP}: Learning guided Quality Cut selection for faster Technology Mapping},
  year      = {2025},
  isbn      = {9798400710773},
  doi       = {10.1145/3676536.3676797},
  booktitle = iccad,
  articleno = {169},
  numpages  = {6}
}

@inproceedings{pan2025pigmap,
  author    = {Pan, Hongyang and Lan, Cunqing and Liu, Yiting and Wang, Zhiang and Shang, Li and Zeng, Xuan and Yang, Fan and Zhu, Keren},
  title     = {Physically Aware Synthesis Revisited: Guiding Technology Mapping with Primitive Logic Gate Placement},
  year      = {2025},
  isbn      = {9798400710773},
  doi       = {10.1145/3676536.3676676},
  booktitle = iccad,
  articleno = {171},
  numpages  = {9}
}

@inproceedings{calvino2023emap,
  author    = {Calvino, Alessandro Tempia and De Micheli, Giovanni},
  booktitle = iccad,
  title     = {Technology Mapping Using Multi-Output Library Cells},
  year      = {2023},
  volume    = {},
  number    = {},
  pages     = {1-9},
  doi       = {10.1109/ICCAD57390.2023.10323999}
}

@article{shang2025dualoutput,
  author   = {Shang, Liuting and Lu, Sheng and Jung, Sungyong and Liang, Qilian and Pan, Chenyun},
  journal  = tcad,
  title    = {Novel FPGA Technology Mapping for Dual-Output LUTs: Methodology and Application},
  year     = {2025},
  volume   = {},
  number   = {},
  pages    = {1-1},
  doi      = {10.1109/TCAD.2025.3616069}
}

@article{liu2025maptune,
  author   = {Liu, Mingju and Robinson, Daniel and Li, Yingjie and Maximilian Kuehn, Johannes and Liang, Rongjian and Ren, Haoxing and Yu, Cunxi},
  title    = {{MapTune}: Versatile {ASIC} Technology Mapping via Reinforcement Learning Guided Library Tuning},
  year     = {2025},
  issn     = {1084-4309},
  doi      = {10.1145/3748507},
  journal  = todaes
}

@article{fu2025temacle,
  author   = {Fu, Rongliang and Wang, Chao and Yu, Bei and Ho, Tsung-Yi},
  journal  = tcad,
  title    = {{TeMACLE}: A Technology Mapping-Aware Area-Efficient Standard Cell Library Extension Framework},
  year     = {2025},
  volume   = {44},
  number   = {8},
  pages    = {3034-3045},
  doi      = {10.1109/TCAD.2025.3529802}
}

@inproceedings{zheng2025chatls,
  author    = {Zheng, Haisheng and Wu, Haoyuan and He, Zhuolun},
  booktitle = dac,
  title     = {{ChatLS}: Multimodal Retrieval-Augmented Generation and Chain-of-Thought for Logic Synthesis Script Customization},
  year      = {2025},
  volume    = {},
  number    = {},
  pages     = {1-7},
  doi       = {10.1109/DAC63849.2025.11132969}
}

@inproceedings{huang2025llsm,
  author    = {Huang, Shan and Li, Jinhao and Yu, Zhen and Ye, Jiancai and Xu, Jiaming and Xu, Ningyi and Dai, Guohao},
  title     = {{LLSM}: {LLM}-enhanced Logic Synthesis Model with {EDA}-guided {CoT} Prompting, Hybrid Embedding and {AIG}-tailored Acceleration},
  year      = {2025},
  isbn      = {9798400706356},
  booktitle = aspdac,
  pages     = {974–980},
  numpages  = {7}
}

@article{AIG,
  author   = {Hellerman, Leo},
  journal  = {IEEE Transactions on Electronic Computers},
  title    = {A Catalog of Three-Variable Or-Invert and And-Invert Logical Circuits},
  year     = {1963},
  volume   = {EC-12},
  number   = {3},
  pages    = {198-223},
  doi      = {10.1109/PGEC.1963.263531}
}

@misc{ABC,
  title        = {{ABC}: A System for Sequential Logic Synthesis and Verification},
  author       = {{Berkeley Logic Synthesis and Verification Group}},
  year         = {Version 1.01},
  howpublished = {http://www.eecs.berkeley.edu/~alanmi/abc/}
}

@misc{mockturtle,
  title        = {mockturtle: A {C++} Logic Network Library},
  author       = {{EPFL Integrated Systems Laboratory}},
  year         = {Accessed on November 2025},
  howpublished = {https://github.com/lsils/mockturtle}
}

@misc{gpt5,
  title  = {Introducing GPT-5},
  author = {OpenAI},
  year   = {2025},
  url    = {https://openai.com/index/introducing-gpt-5}
}

@misc{deepseekv3,
  title         = {{DeepSeek-V3} Technical Report},
  author        = {DeepSeek-AI},
  year          = {2024},
  eprint        = {2412.19437},
  archiveprefix = {arXiv},
  primaryclass  = {cs.CL},
  url           = {https://arxiv.org/abs/2412.19437}
}

@misc{qwen3,
  title         = {Qwen3 Technical Report},
  author        = {Qwen Team},
  year          = {2025},
  eprint        = {2505.09388},
  archiveprefix = {arXiv},
  primaryclass  = {cs.CL},
  url           = {https://arxiv.org/abs/2505.09388}
}

@misc{openevolve,
  title     = {{OpenEvolve}: an open-source evolutionary coding agent},
  author    = {Asankhaya Sharma},
  year      = {2025},
  publisher = {GitHub},
  url       = {https://github.com/algorithmicsuperintelligence/openevolve}
}

@misc{alphaevolve,
  title         = {{AlphaEvolve}: A coding agent for scientific and algorithmic discovery},
  author        = {Alexander Novikov and Ngân Vũ and Marvin Eisenberger and Emilien Dupont and Po-Sen Huang and Adam Zsolt Wagner and Sergey Shirobokov and Borislav Kozlovskii and Francisco J. R. Ruiz and Abbas Mehrabian and M. Pawan Kumar and Abigail See and Swarat Chaudhuri and George Holland and Alex Davies and Sebastian Nowozin and Pushmeet Kohli and Matej Balog},
  year          = {2025},
  eprint        = {2506.13131},
  archiveprefix = {arXiv},
  primaryclass  = {cs.AI},
  url           = {https://arxiv.org/abs/2506.13131}
}

@article{fu2026chop,
  author   = {Fu, Rongliang and Zhang, Ran and Zheng, Ziyang and Shi, Zhengyuan and Pu, Yuan and Huang, Junying and Yu, Bei and Xu, Qiang and Ho, Tsung-Yi},
  journal  = tcad,
  title    = {{CHOP}: Clustered Hybrid Optimization for Logic Synthesis with Self-Supervised Prediction},
  year     = {2026},
  volume   = {},
  number   = {},
  pages    = {},
  doi      = {10.1109/TCAD.2026.3650860}
}

@article{lee2023peephole,
  author   = {Lee, Siang-Yun and Micheli, Giovanni De},
  journal  = tcad,
  title    = {Heuristic Logic Resynthesis Algorithms at the Core of Peephole Optimization},
  year     = {2023},
  volume   = {42},
  number   = {11},
  pages    = {3958-3971},
  doi      = {10.1109/TCAD.2023.3256341}
}

@inproceedings{dc_rewriting,
  author    = {Calvino, Alessandro Tempia and De Micheli, Giovanni},
  booktitle = date,
  title     = {Scalable Logic Rewriting Using Don't Cares},
  year      = {2024},
  volume    = {},
  number    = {},
  pages     = {1-6},
  doi       = {10.23919/DATE58400.2024.10546807}
}

@inproceedings{fu2026dclog,
  author    = {Fu, Rongliang and Shen, Libo and Wang, Ziyi and Lei, Zhengxing and Wang, Zixiao and Huang, Junying and Yu, Bei and Ho, Tsung-Yi},
  booktitle = aspdac,
  title     = {{DCLOG}: Don't Cares-based Logic Optimization using Pre-training Graph Neural Networks},
  year      = {2026},
  volume    = {},
  number    = {},
  pages     = {793-799},
  doi       = {10.1109/ASP-DAC66049.2026.11420303}
}

@inproceedings{fu2026elogic,
  author    = {Fu, Rongliang and Xuan, Wei and Yin, Shuo and Hu, Guangyu and Chen, Chen and Zhang, Hongce and Yu, Bei and Ho, Tsung-Yi},
  booktitle = date,
  title     = {{eLogic}: A E-Graph-based Logic Rewriting Framework for Majority-Inverter Graphs},
  year      = {2026},
  volume    = {},
  number    = {},
  pages     = {1-6},
  doi       = {}
}

@inproceedings{calvino2022mapping,
  author    = {Calvino, Alessandro Tempia and Riener, Heinz and Rai, Shubham and Kumar, Akash and De Micheli, Giovanni},
  booktitle = aspdac,
  title     = {A Versatile Mapping Approach for Technology Mapping and Graph Optimization},
  year      = {2022},
  volume    = {},
  number    = {},
  pages     = {410-416},
  doi       = {10.1109/ASP-DAC52403.2022.9712552}
}

\end{document}